# *The Order of Physical Law*

Ted Sichelman[†]

## Abstract

First-order legal relations specify the duties of legal actors. For instance, the duty not to trespass derives from a first-order law. Second-order legal relations generally concern the intentional, volitional acts of legal actors exercising legal powers to change first-order laws or legal relations. For example, a land owner may exercise a second-order power to change another legal actor's duty not to trespass to a legal permission to enter the owner's land. This article adapts the notion of legal order to propose a theory of first- and higher-order physical laws, contending that current physical theories implicitly (and wrongly) assume that essentially all physical processes can be modeled using first-order laws. Incorporating second- and higher-order structures from legal models into physical theories provides a novel approach for framing problems in physics, such as the process of quantum measurement. Specifically, quantum measurement is better explained as a fundamentally second-order physical process, which alters the underlying first-order physical "microlaws" governing the evolution of the quantum system.

## 1. Introduction

*Physical laws*, such as Newton's First Law, Einstein's gravitational field equations, and the various equations governing the evolution of quantum states are typically characterized as general rules describing how fundamental physical objects in the world interact and change in space and time, subject to certain external conditions and constraints (Feynman, 1967). *Social laws*, such as laws regarding theft, taxes, and speeding, are often characterized as human-made rules of how legal actors should behave, the violation of which results in liability, i.e., exposure to punishment and other sanctions (Austin, 1832; Corbin, 1924).

Legal theorists have long recognized that social laws have multiple "orders." First-order laws or legal relations concern obligations (Sumner, 1987)—that is, rules that require a legal actor to engage (or not to engage) in some specified action (sometimes, more broadly, rules that require some state of the world to entail or not). Second-order laws or legal relations characterize legal powers to change, terminate, or create first-order laws or legal relations (Weiner, 2021). For instance, two private parties may contract to impose obligations on one another to perform specific actions (e.g., for A to deliver 1000 widgets by Thursday in return for a $500 payment from B) that did not exist prior to the contract. In essence, a contract imposes a "microlaw"— namely, a legal rule that applies to a small number of legal actors, which in turn governs the behavior of those actors. Similarly, a legislature may exercise its power to pass oridinary laws that impose obligations on multitudes of legal actors and a court may use its power to decide a dispute between parties in front of it, potentially imposing a payment or other obligations on the defendant (again, an effective microlaw). Third-order laws or relations concern the change, terminate, or create second-order laws or legal relations (Weiner, 2021). And so forth.

---

[†] Professor or Law, University of San Diego, 5998 Alcala Park, San Diego, CA 92130, USA, tsichelman@sandiego.edu. J.D., Harvard Law School; M.S., Physics, Florida State University; A.B., Philosophy (Program in the History and Philosophy of Science), Stanford University.

(Importantly, note that while "order" in legal systems is related to "order" in logic, the concepts differ in important respects, discussed further in Part 2. And the legal concept differs even more from notions of "order" in differential equations and related mathematical fields.)

Although social laws and physical laws concern different subject matter, extending a centuries-old lineage of thought, this article asserts that they both exhibit a similar underlying structure, particularly the legal notion of the "order" of laws. Unlike legal theorists, physicists and philosophers of science have yet to classify physical laws as first-order, second-order, and so forth. In this sense, physicists and philosophers nearly universally assume that physical laws, whether viewed merely regularities in nature or laws proper, are first-order and unchanging. As such, there is very little of a scientific nature that attempts to explain the genesis, change, and termination of physical laws (Ross, Ladyman, and Kincaid, 2013). And the small number of exceptions that address how physical law might itself evolve do so either by positing additional first-order dynamics or by making conceptual reference to evolving "meta-laws," without offering formal second-order representations or operators (e.g., Smolin and Unger, 2015). As such, these treatments are wanting about why and how physical laws may change, much less the origin and termination of physical laws.

Here, I contend that not only is it feasible to categorize physical law according to orders similar to those used in legal models, but in so doing, some unresolved issues in physics—such as quantum measurement—can be recast in a manner that renders them more amenable to resolution. As an example, this article proposes that the act of quantum measurement involves in essential part a second-order physical process. Like a judge deciding between whether a defendant is liable or not, the process of quantum measurement acts to select a final state from possible states of the world via a physical process a critical portion of which is wholly independent of ordinary (i.e., first-order) physical processes and laws. In contrast to first-order physical processes, which merely change the first-order physical *state* of a system, but not the *rules* that the system must "obey," the second-order process of quantum measurement in effect selects one classical-like rule from a superposition of possible classical-like rules, the result of which dictates the measured state of the system.

Such an approach dispels with several critical conundrums surrounding quantum measurement. It avoids the issue that the measurement device itself is quantum and, as such, there appears to be no clear mechanism to collapse the joint wave function of quantum system and device. Continuing the legal analogy, although a judge is certainly subject to first-order rules—e.g., the judge is prevented by ordinary criminal and tort laws from stepping down from the bench and striking a disputatious attorney—the judge's judgment power per se is entirely separate from these ordinary first-order rules. In other words, judgment sounds in power and limitations on power, rather than turning on first-order rights (in the strict sense) and obligations. Similarly, measurement turns on second-order processes akin to the exercise of legal powers, rather than the first-order, ordinary physical laws that govern the evolution of quantum systems.

To be certain, the approach offered here leaves open many questions. Most importantly, from a formal mathematical perspective, how does the second-order aspect of quantum measurement operate and what precisely triggers the process? For instance, what mathematical theory explains how the microlaws that govern specific quantum states are selected? And are there testable predictions of this approach that distinguish it from other approaches? While this article briefly addresses these questions, a more exhaustive treatment will be proposed in future work (Sichelman, 2025). In contrast, this article aims to provide a conceptual framework for the model, offering a foundation for further exploration.

More specifically, this article contributes to the literature in at least three important respects. First, it extends theories of social, "juridical" law to offer a richer explanatory account of physical law, including a description of the structure of—and effective changes in—physical law. Second, it applies this account to the problem of quantum measurement, making the novel assertion that measurement is, in significant part, a second-order physical process. Third, beyond helping to explain quantum measurement, such an approach may assist in answering even deeper questions about the nature of physical law in a rigorous mathematical manner.

Section 2 provides a brief description of how social laws like property, contract, and constitutional law can be modeled in terms of orders of legal relations. Relying on the landmark treatment of legal relations set forth by Hohfeld (1913), it further describes efforts by legal philosophers to formalize these legal relations into logical and mathematical systems, including deontic logic. Finally, Section 2 posits extensions of the standard formalizations to probabilistic interpretations of legal relations. Section 3 extends these formalizations to describe how scientific law may be modeled in terms of orders of deontic-modal relations.[1] Section 4 briefly concludes.

## 2. The Orders of Social Law

To fully understand the notion of a second-order, *physical* system, it is instructive to explore the nature of a second-order, *legal* system. Indeed, as Corbin (1921) recognized, "[r]ules of physics and rules of law are alike in that they enable us to predict physical consequences and to regulate our actions accordingly." Unlike physics, however, legal theory has long provided a formal framework for understanding not just rules that regulate behavior, but also secondary rules, or "rules about rules" (Hart, 1961), which describe how laws arise, change, and expire. This article posits that understanding the nature of secondary rules in law aids in a deeper understanding of how laws operate in physical theories.

Beginning at least with the foundational work of Wesley Hohfeld (1913), legal theorists have proposed that the law exhibits a formal logical structure. Hohfeld (1913) posited that there are eight logically related "fundamental legal relations" that can be combined to describe all legal phenomena. Later scholars (e.g., Allen, 1974; Allen & Saxon, 1997; Kanger, 1972; Lindahl, 1977) formalized these relations using variants of deontic logic. Extending the Hohfeldian-deontic formalism to account for probabilistic legal relations offers a foundation for applying formal models of legal systems to stochastic physical theories.

2.1 First-Order Legal Relations

The starting point for these formalizations is the set of *first-order* relations identified by Hohfeld (1913), namely *duty*, *privilege*, *right* (in the strict sense), and *no-right*.[2] As an illustrative example, assume that legal actor *A* owns a piece of land *L*. *B* is some other legal actor that has no ownership interest in *L*. One standard "incident" of real property ownership is the "right to exclude," that is, the right of the owner to prevent trespassers (barring limited exceptions) from entering the owner's land. In other words, if the non-owner, *B*, has no valid excuse to enter *L*, in Hohfeldian terms, the owner *A* has a "right"—specifically, vis-à-vis the third party, *B*—that *B* not enter *L*. Because a Hohfeldian *right* is a precise form of legal right, it is useful to refer to it as a "strict-right" (Sichelman, 2022, 2024). In turn, A's *strict-right* implies that *B* has a "correlative" Hohfeldian "duty *not* to enter *L*. Just as *A* having a *strict-right* implies that *B* has a *duty*, if *A* has no *right* (termed a *no-right* by Hohfeld (1913)), then correlatively *B* has no *duty* (termed a *privilege* by Hohfeld (1913)).

---
[1] Although this extension is related to modal theories of quantum mechanics (e.g., Van Fraassen (1981)), the approach here differs in several notable respects (Sichelman, 2025).
[2] Hohfeld (1913) is notoriously difficult to parse, even for legal theorists. An edited and annotated version of Hohfeld (1913), which clearly explains the Hohfeldian typology as well as common misconceptions regarding it, can be found in Sichelman (2022).

Formally, the first-order Hohfeldian relations can be defined in terms of one another via operators.[3] Let a *strict-right* be symbolized by the letter *r*. In this event, a *no-right* is just ~*r* (where "~" indicates negation).[4] Using these abbreviations, *A*'s *strict-right* vis-à-vis *B* that *B* not enter *L* may be written as:

$$A_r B \text{ (B not enter L)} \qquad (1)$$

In general, all forms of "complete," classical first-order legal propositions (i.e., statements like (1)) concern three elements. First, there must be at least two legal actors to which the relation pertains. These actors may be real persons or artificial entities, such as corporations, partnerships, and even the government ("the State"). For instance, in (1), *A* and *B* are the two legal actors of concern. Second, there must be a "specific" first-order legal relation between the two actors. For first-order classical relations, either the two actors have a *strict-right/duty* relation or a *no-right/privilege* relation. That is, one actor (here, *B*) either owes a *duty* to another actor (here, *A*) or not, such that A holds a strict-right (*r*) or not (~*r*). Third, the specific relation between the two actors will concern an action that one of the actors (or a third party) performs (or not) or, more generally, specific "states of affairs" of the world (that obtain or not). Typically, the state of affairs is an action that the actor with a *duty* or *privilege* must engage in (a "positive" *duty*), must not engage in (a "negative" *duty*), may engage in (a "positive" *privilege*), or may forgo (a "negative" *privilege*).

As such, the general form of any complete first-order legal proposition is as follows:

$$[\text{Actor \#1}]_{[\textit{legal relation}]}[\text{Actor \#2}] \text{ (a state of affairs that the legal relation concerns)} \qquad (2)$$

Letting "Actor #1" be *X*, the first-order jural (i.e., legal) relation be $j_1$,[5] Actor #2 be *Y*, and the state of affairs, *S*, then all first-order complete classical jural propositions, $J_1$, take the following form:

$$J_1 = X_{j_1} Y(S) \qquad (3)$$

In view of the "correlativity" principle of the Hohfeldian framework, by convention, it is possible to orient this form always in "*strict-right* notation," meaning that *X* is always the legal actor that holds a *strict-right* (or not) and *Y* is always the legal actor that is subject to a *duty* (or not). In this event:

$$j_1 = r \text{ or } \sim r \tag{4}$$

---

[3] A more detailed description of the formalization of the Hohfeldian typology presented in Sections 2.1-2.4 can be found in Sichelman (2024).
[4] Hohfeld (1913) termed a *strict-right* and *no-right* as "opposites." In more precise logical terms, ~*r* is the negation (or absence) of *r*.
[5] I use "j" instead of the letter "l" to denote legal relations for two reasons: one, it is easier to distinguish (particularly in lower-case) when it is adjacent to the number "1"; two, it follows Hohfeld's original terminology of "jural relation."

Thus, in *strict-right* notation, first-order legal propositions must always take the form:

$$J_1 = X_{j_1} Y(S) \text{ where } j_1 = r \text{ or } \sim r \tag{5}$$

The application of any precise legal rule concerning a first-order relation involving two particular legal actors and a specific state of affairs may be expressed in the form of (5).[6]

The Hohfeldian formalization anticipates the more rigorous formalization of first-order legal norms via deontic logic (Mally, 1926; von Wright, 1951). The traditional definitional scheme of deontic logic centers around the *obligation*, which is fully equivalent in definitional terms to the Hohfeldian duty. In deontic logic, the obligation is typically taken to be primitive, with other conceptions defined in terms of it. Specifically,

$PEp =_{def} \neg OB \neg p$ (a state of the world is permissible iff, i.e., if and only if, it is not obligatory that the state the world not obtain) (6)

$IMp =_{def} OB \neg p$ (a state of the world is impermissible iff it is obligatory that the state of the world not obtain) (7)

$OMp =_{def} \neg OBp$ (a state of the world is omissible iff it is not obligatory that it obtain) (8)

$OPp =_{def} (\neg OBp \& \neg OB \neg p)$ (a state of the world is optional iff it is both permissible and omissible) (9)

where OB is obligation, PE is permissibility, IM is impermissibility, OM is omissibility, and OP is optionality and p is some action performed by an actor subject to the deontic relations or, more generally, p is some general state of affairs (i.e., state of the world) that concerns the actor.

It is readily apparent that if one equates a positive Hohfeldian duty with the deontic obligation, then the definitional scheme is identical.[7] This allows the Hohfeldian scheme to be formalized similarly to standard deontic logic or alternative deontic logics, for which there is a rich literature of formalizations (e.g., Sergot, 2001; McNamara, 2006). Because deontic logic is a variant of modal logic, which plays an important role in formalizing physical law, the formal structure of the first-order Hohfeldian relations can readily be adapted to model first-order physical law (see Section 3).

2.2 Second- and Higher-Order Legal Relations

However, because standard deontic logic is, in the Hohfeldian sense, only first-order, there is a complete absence of a connection between standard deontic logic and the Hohfeldian

---

[6] By "precise," I mean that the legal rule—when applied to a state of affairs (i.e., a set of facts)—yields a unique answer (i.e., whether a *right* or *no-right* exists between the given legal actors). I relax this assumption in Section 4.

[7] Specifically, in this instance, a Hohfeldian positive privilege is equivalent to a standard deontic permissibility, the negation of a Hohfeldian positive privilege is equivalent to a deontic impermissibility; and the negation of a Hohfeldian positive duty is equivalent to deontic omissibility. Additionally, the coupling of a Hohfeldian positive privilege and the negation of a Hohfeldian positive duty is a deontic option.

second- and higher-order relations, which concern the change, creation, and termination of lower-order legal relations. In other words, any change in legal rules must emanate from outside the system of standard deontic logic (cf. Sergot, 2001; Dong and Roy, 2017). Standard modal logic operates in essentially the same manner (van Bentham, van Eijck & Stebletsova, 1994).

Hohfeld (1913) described second-order legal *power* relations that describe (from within the formal system) a legal actor's ability to create, change, or extinguish lower-order legal relations. Referring back to our example of a landowner, suppose now that A provides permission to B to enter the land L. Formally, the permission is effectuated through A's exercise of a *legal power*. Specifically, when A exercises pwer, the *strict-right* relation, $A_rB$ *(B not enter L)*, will transform to a *no-right* relation, $A_{\sim r}B$ *(B not enter L)*. A loses a right to keep B off the land and B gains a privilege to enter A's land.

Importantly, a second-order power does not simply effectuate a change in a *legal state*, but a change in the *legal relation* that applies to the legal state, which in turn may affect the legal state. Following our example, B was originally under a duty A not to enter A's land. If B unlawfully enters A's land, then the legal state of concern changes from no-violation to one of violation. This change in B's legal state is caused by the *physical* action undertaken by B. In general, changes in legal states often occur when a legal actor's physical or mental states materially change, but these changes typically do not change the underlying legal rule at issue—here, B's duty not to trespass. Thus, in contrast to ordinary first-order changes in legal *states*, second-order changes actively alter the legal *rule* that applies to a given state.[8]

In this regard, it is important to note that standard second- and higher-order *logics*—although related to second- and higher-order characterizations of *legal relations*—do not adequately provide a formal language to describe the change, creation, and termination of legal rules (or physical laws). Specifically, standard higher-order logics essentially allow for quantification over predicates (properties and propositions) and sets, in addition to quantification over individual elements within sets, as in first-order logic (Shapiro, 1991). But this additional quantification contains no primitive dynamic operators of the sort in Hohfeldian second-order relations. Even the more recent development of dynamic and temporal logic, which describe how propositions change across time or states, generally do not formalize the mechanisms by which such changes occurs (Harel, Kozen & Tiuryn, 2000; Stirling, 1992). Rather, only Hohfeldian extensions of dynamic and temporal logic, such as the one proposed by Dong and Roy (2017), aim to formalize the mechanisms driving such changes.

2.3 Probabilistic Legal Relations

Another limitation of standard deontic logic as well as standard Hohfeldian approaches is their inability to deal with probabilistic legal relations (Sichelman, 2024). Legal "indeterminacy" can either be *epistemological* or *ontological* in nature. Epistemological indeterminacy applies to situations in which there is a "correct" underlying legal relation that applies to the situation at hand but legal observers lack knowledge regarding the legal relation. For example, at the time a lawsuit is filed, parties may be unaware of key facts essential to determining liability, which prevents them from predicting the outcome of the suit. If the law, when fully applied to a complete set of facts, produces a definite result, the indeterminacy is purely epistemological.

---

[8] There are contingent legal duties and rights, which are triggered only upon some occurrence of a particular mental or physical action or condition. For instance, the "right to self-defense" is only contingent upon some other legal actor's actions triggering such a "right" (actually, a Hohfeldian privilege). Because contingent legal relations can be known in advance, they are not changes in the law via a second-order power, but instead, wholly specifiable with first-order relations subject to defeasible logic (cf. Sichelman, 2022).

It was not until the 20th century that "inherent," ontological legal indeterminacy fully took root. On this approach, at least for certain classes of cases and legal circumstances, even complete knowledge of the legally relevant facts and law cannot fully determine the rights and obligations of legal actors. Ontological indeterminacy can arise from "vagueness or indeterminacy of legal doctrine"; "uncertainty as to the impact evidence will have on the decisionmaker"; idiosyncratic behavior in enforcement and adjudication; and the influence of unknowable, "extra-legal" factors on the regulatory and judicial process (Yablon, 1996).[9] Notably, the "post-classical," ontological approach to legal indeterminacy contrasts with the "classical," formalist approach, which asserts all forms of legal indeterminacy are epistemic in nature, in a manner that is similar to the contrast between indeterminacy in quantum and classical mechanics.

When a court issues a judgment, it uses its judicial "power" to resolve ontological indeterminacy by specifically determining the rights and obligations of the applicable legal actors (Kennedy, 1973). From a Hohfeldian perspective, judicial power is the same form of power—albeit in a more limited form—as that exercised by legislatures. In general, the adjudicatory judicial power—which is quite separate from the legislative-like, "common law" power enjoyed by Anglo-American courts to actually fashion law itself—is a limited form of power that provides courts the ability to "collapse" probabilistic legal relations into derminate, classical legal relations upon judgment.[10]

For example, suppose a patentholder files a lawsuit accusing a large technology company of infringing its patent. Prior to any judgment, expert legal observers agree that the courts will find the patent invalid with a 40% chance and valid with a 60% chance. Such indeterminacy in the legal rule that applies is, at least on the post-classical approach, wholly ontological. Upon the final judgment, there is a certain outcome—either the patentholder win (rule #1) or not (rule #1), but not a probabilistic "superposition" of the two. Yet, prior to judgment, parties will act as if the applicable legal rule is a superposition (here, 40/60) of the two possible legal rules ("microlaws") that could apply post-judgment. After judgment, all legal observers will agree that the infringer is either under a duty not to infringe *or* has a privilege to infringe. Unlike the state of the world prior to judgment, there is no middle ground following judgment. What was once ontological indeterminacy is eliminated by the power of the court. The court's ability to bring certainty to bear ultimately stems from a second-order process in which it decides a pending case, rather than merely from the court's or parties' first-order duties and rights.

---

[9] One might counter that the legal relations would be knowable if sufficient information about the extra-legal factors determining enforcement, adjudication, and jury decisionmaking could be obtained. As an initial matter, to the extent human decisions are themselves ontologically indeterminate, it would be impossible in principle to make perfect predictions. Even still, *legal* ontological indeterminacy specifically concerns whether there is residual indeterminacy even after all of the *legally* relevant facts and applicable law are fully known. To be certain, "applicable law" may take into account factors beyond mere interpretation, such as economic, institutional, social, and other effects, as well as general "fairness," but even then, legal scholars generally believe that legal ontological indeterminacy remains, at least in "hard" cases (cf. Capps, 2025).

[10] Even when an outcome is completely certain, when a court issues a judgment—for instance, for the defendant to pay the plaintiff damages—it is also exercising a power. Here, the discussion of "adjudicatory" power solely concerns the court's power *to decide* the legal rights and obligations of the parties where those rights and obligations are indeterminate.

### 2.4 Mathematically Formalizing the Legal Relations and the Notion of Judgment

To fully appreciate the insights that the notion of "order" in legal relations can offer to an understanding of physical law, it is useful to examine a more mathematical formulation of the legal relations (Sichelman, 2024). In contrast to the purely logical formalizations of the Hohfeldian and related deontic relations, the mathematical formulation relies on a tensor-based approach that draws upon information theory.

#### 2.4.1 A Formalization of the Classical Legal Relations

For classical first-order relations, an actor either has a strict-right (or not) corresponding to the duty (or absence of duty) of another actor with respect to some action or state of the world (e.g., a certain duty not to trespass). In this event, we can represent the duty by a classical bit being "on." If the duty is eliminated, again for example by A providing permission to B to enter the land, then the classical bit is "off."

Because these classical bits will become quantum (or at least quantum-like) bits in the post-classical formalization, instead of using a scalar bit to represent classical first-order relations, it is more instructive to use vectors. Specifically, recall from (5) above, we can represent a first-order legal proposition as follows:

$$J_1 = X_{j_1} Y(S) \text{ where } j_1 = r \text{ or } \sim r \qquad (10)$$

Here $J_1$ represents a first-order jural (legal) proposition, X is the right-holder (or not), Y is subject to a duty (or not), S is the state of affairs applicable to the relation, and $j_1$ is the jural (legal) relation. This jural relation in a classical system is either a strict-right (and corresponding duty) or a no-right (and corresponding lack of duty). Thus, in vector formalism, we can represent a strict-right as an "on" vector and a no-right as an "off" vector:

$$r_1 = \begin{pmatrix} 1 \\ 0 \end{pmatrix} \text{ and } \sim r_1 = \begin{pmatrix} 0 \\ 1 \end{pmatrix} \quad \text{(matrix notation)} \qquad (11)$$

Importantly, these vectors do not represent the underlying state of the legal system per se, but rather whether the applicable legal actors are subject to a certain duty-based legal rule (or not). In essence, states of the world subject to an "on" vector are legally required and states of the world subject to an "off" vector are not legally required. In our example, when the duty is "on," the state of the world that is required is one where B is not located on A's land. We shall see in the next section that the physical laws have a very similar structure.

Recall that a second-order power relation changes the underlying first-order relations. Thus, when the landowner A provides permission to B to enter A's land via a legal power, the "on" bit of the applicable duty changes to an "off" bit. In general, a second-order legal proposition written in *power* notation must either contain a *power* (which changes a first-order relation) or a the absence of power (a Hohfeldian *disability*, which leaves intact a first-order relation) (Hohfeld, 1913; Sichelman, 2024):

$$J_2 = (X_{j_2}Y(S_2))(J_1) \text{ where } j_2 = r_2 \text{ or } \sim r_2 \qquad (12)$$

Here, $J_2$ is a second-order legal proposition that represents the operation of a second-order legal power (or not) on a first-order legal proposition, $J_1$. If a second-order legal relation $j_2 = r_2$ (i.e., $j_2$ is a *power*), then $X$'s exercise of the *power* will transform the specific legal relation of $J_1$ (i.e., $j_1$) into its negation. So if $j_1$ is a *strict-right* (i.e., the vector $(1,0)$), the operation of a *power* on $j_1$ will transform the *strict-right* into a *no-right* (i.e., the vector $(0,1)$). Thus, $j_2$ can be thought of as a mathematical operator that either transforms a first-order vector relation into its negation (i.e., in the event $j_2$ is a *power* ($r_2$)) or leaves a first-order vector relation intact (i.e., in the event $j_2$ is a *disability* ($\sim r_2$)).

In this regard, one can represent a second-order *power* ($r_2$) by a second-rank bit flip tensor, and a second-order *disability* ($\sim r_2$) by a second-rank identity tensor. In matrix notation, $r_2$ and $\sim r_2$ are represented by the following forms:

$$r_2 = \begin{pmatrix} 0 & 1 \\ 1 & 0 \end{pmatrix} \text{ and } \sim r_2 = \begin{pmatrix} 1 & 0 \\ 0 & 1 \end{pmatrix} \qquad (13)$$

Third and higher-order classical powers also flip or leave intact the lower-order relations and thus may be represented by the same second-rank bit-flip and identity tensors (Sichelman, 2024).

When a legal actor exercises a legal power, whether in a classical or post-classical system, from a first-order perspective, all of the legal states altered by the power automatically and instaneously update. However, like strict-rights and duties, there are notable differences in classical and post-classical systems when powers are exercised.

### 2.4.2 A Formalization of the Post-Classical Relations

Recall that the key shift from classical to post-classical relations is the view that legal relations can be *ontologically* indeterminate. Even with all of the relevant law and facts in hand, expert legal observers cannot do better than to assign odds to the potential outcomes of a final judgment in court (cf. Holmes, 1897). Such ontological indeterminacy is analogous to the indeterminacy of pure states prior to measurement in quantum mechanics (Sichelman, 2024; Godfrey, 2024).

At least according to the dominant Copenhagen interpretation, a pure quantum *physical* state exists in an ontologically indeterminate superposition, such that there is no particular classical-like state (an "observable") of the quantum system prior to measurement. Rather, it is only upon measurement that the system "collapses" to a particular measurable state. For ontologically indeterminate *legal* states, there is similarly no "measurable" legal state prior to judgment but a superposition that only collapses to a measurable, "classical" (in the Hohfeldian sense) state upon judgment. (Notably, while post-classical legal indeterminacy is ontological, the pre-measurement legal state of superposition is not "ontic" in the sense of being realizable in the form of a measurement, i.e., a legal judgment—an aspect that provides an important insight into the nature of quantum states.)

Following this analogy, one can posit that a quantum first-order legal proposition is of the following form:

$$|J\rangle_1 = X_{|j\rangle_1} Y(S_1) \qquad (14)$$

where
$$|j\rangle_1 = a\,|j_r\rangle + b\,|j_{\sim r}\rangle \qquad (15)$$

Here, the classical legal proposition $J_1$ is replaced with the "quantum" legal proposition, $|J\rangle_1$, where the applicable actors and state of the world are the same as in the classical case, but the relevant legal relation is now a quantum-like superposition of the right and no-right states, reflected by $|j_r\rangle$ and $|j_{\sim r}\rangle$. As before, $|j_r\rangle$ can be represented by the vector (1 0) and $|j_{\sim r}\rangle$ by the vector (0 1), making the first-order post-classical legal relations much like the superposition of quantum spin states around the z-axis.

However, there is one key difference between quantum physical states and ontologically indeterminate legal relations—the presence of self-interference in pure quantum states, as illustrated by the double-slit experiment. In this experiment, the wave-like nature of the probabilities governing the "location" of particles, such as electrons, prior to measurement causes constructive and destructive self-interference among the probability waves that is wholly non-classical in nature. In general, post-classical legal relations do not exhibit such self-interference.[11]

As such, the probabilities in (15), unlike the physical quantum state, can simply be represented by real numbers, so that the probability of measuring a strict-right state is $a$ and the probability of measuring a no-right state is $b$ (which must equal $1-a$). This formulation of the "quantum" legal relation resembles irreducible, diagonalized "mixed" quantum states and classical probabilities, rather than pure states and associated quantum probabilities, which depend on the *square* of the factors $a$ and $b$. As such, one might argue that the post-classical legal state aligns more closely with "classical" rather than "quantum" properties, at least from the perspective of physics.

Nonetheless, the absence of interference properties does not imply that indeterminate legal states are somehow classical in the sense of exhibiting only epistemic indeterminacy. Notably, ontological indeterminacy in law does not turn on the presence of self- or any other kind of interference in legal systems (Sichelman, 2024; Godfrey, 2024). Here, I argue by analogy to legal systems that ontological indeterminacy in physical systems does not depend upon interference phenomena. Although this view may seem contrary to prevailing interpretations, it aligns with the widespread view that merely diagonalizing the density matrix of a quantum state does not yield a quantum measurement.

More specifically, when diagonalized density matrices merely reflect the lack of knowledge of observers of which specific pure state describes a physical system, then the underlying probabilities are epistemic as in classical physics. But when the diagonalization results from decoherence that eliminates the self-interference of an otherwise pure state, ontological indeterminacy typically remains. In such cases, the residual probabilities may share the same mathematical form as classical probabilities, but they are not classical in the sense that they merely arise from ignorance of an otherwise determinate system.

---

[11] Notwithstanding, quantum-like interference has proven useful in modeling aspects of law, such as the level of formalism in a particular judicial interpretation (Godfrey, 2024). One can also imagine that the competing arguments in favor and against a particular judgment (outcome) "interfere" with one another to create dynamic probability updates, which could be modeled by interfering probability distributions. Yet, such models in law appear to be less of a necessity than in quantum mechanics. Regardless, the key point here is that even if interference is not present in a given rule-based system, *ontological* indeterminacy may exist.

From a mathematical standpoint, classical probability theory can represent either epistemic or ontological indeterminacy. The mere fact that epistemic indeterminacy in classical physics can be modeled using classical probability theory does not imply that all uses of classical probability theory in physics reflect epistemic indeterminacy. Thus, nothing precludes diagonalized mixed states from exhibiting ontological indeterminacy. In cases such as decoherence, some additional operation—i.e., a non-classical measurement—must occur to eliminate the residual indeterminacy. As in legal systems, classical, "first-order" physical measurement can only "reveal" pre-existing states, but it cannot have any ontological effect. As Section 3 explains, ontology-altering measurements must be second-order in nature and are fundamentally non-classical in the physical sense.

2.5 Law as Coercive vs. Law as Descriptive

One potential preemptive criticism of applying the concept of order in social law to physical law is that there is a fundamental difference between the two types of laws: social law is prescriptive in the sense that law is promulgated while physical law is simply a scientific description of regularities found in nature. As such, whatever we might understand about the process of "power" in the legal domain is not applicable to the physical domain, because the physical domain lacks any sort of prescriptive power, or at least we must be agnostic with respect to any prescriptive notion of physical law (cf. Popper, 1945; Lewis, 1973; Beebee, 2000; Siegel, 2001; Rahmatian, 2017).

There are two key responses. First, the Hohfeldian formalism presented here, including the notion of higher-order legal relations such as powers, does not rest upon a prescriptive view of the law. For instance, suppose a large team of anthropologists visited a society and observed which actions (or, more generally, states of the world) led to punishment and liability and which ones did not, without ever examining or inquiring about prescriptive laws per se. If the anthropologists had a very large amount of time and resources, in theory, they could derive the "laws," in the sense of inferred rules of behavior, simply from observed regularities in punishment and liability. This could be achieved even if punishment and liability were probabilistic in nature. This process would be similar to that of physicists "deriving" laws from repeated observation and model refinement.

Even more, suppose that the regularities in punishment and liability occasionally changed, so that during one period certain actions were permitted by law and during another period they were not. Suppose further that the anthropologists determined that at least some of these changes in laws corresponded to certain actions of the society's leaders, special meetings, voting by members of the society, and so forth. Again, without examining what was written down, the anthropologists could infer that these actions sounded in second-order legal relations concerning power to change law rather than the ordinary first-order legal relations.

Indeed, the legal realism movement notably distinguished "law on the books" from "law in action," arguing that the "real" laws are those that result in punishment and liability rather than what is written down in books (Llewellyn, 1930). From this perspective, the anthropological view of law is more than a mere hypothetical designed to illustrate the possibility of deriving law from the observation, but is a necessary exercise for identifying the law that fundamentally governs human behavior. Regardless of whether one is a legal realist, the key point is that social law can be viewed and modeled—including formally—entirely through a descriptive lens, much like physical law. As such, the supposed differentiation between social and physical law on a prescriptive versus descriptive basis loses much of its force.

Second, simply because physical law has a descriptive origin does not preclude that these laws (or some set of laws of which the known laws are an approximation) are ultimately prescriptive (cf. Emery, 2022). Continuing the analogy, that anthropologists observe social laws through regularities does not preclude a prescriptive set of laws that "cause" the observed regularities. Of course, a major difference between social law and physical law is that the observers generally have access to the prescriptive sources of law in legal systems, whereas the observers of scientific laws only describe empirical regularities. Nonetheless, like our anthropologists without access to legal sources, regularities in physical law structurally appear in two flavors. First-order physical regularities describe the positions and temporal evolution of the positions and related states of fundamental physical objects. Second-order regularities describe the change of the "laws" underlying the first-order regularities, including microlaws, which refer to the application of a more general law to a specific physical system.

All physical theories today assume that there is no means to access the source of physical law. Unlike the hypothetical anthropologists who cannot access the written, prescriptive laws—but can observe the (second-order) changes in first-order regularities resulting from certain social acts—nearly all physical theories explicitly or implicitly assume the even stronger claim that there are no second-order regularities that occur in the universe as we observe it. In the next section, I argue, at least conceptually, that this assumption is misguided—notably, quantum measurement is in critical part a second-order process with a directly observable physical result, namely the so-called collapse of the quantum state.

3. **The Orders of Physical Law**

As with social law, one can formalize the notion of physical law using a Hohfeldian approach. Indeed, deontic logic is commonly viewed as a branch of modal logic, which has been employed widely to provide a formal description of physical law. In another sense, one can envision modal logic as a system of deontic logic in which the "actors" have no free will. Absent free will, deontic obligations become modal necessities and deontic permissions become modal possibilities. As such, we can connect the use, both explicit and implicit, of modal logic to model physical law (e.g., Deutsch & Marletto, 2015; Stern, 1988; Dalla Chiara, 1977) to the use of deontic logic and related models of social law to model physical law.

However, like standard deontic logic, standard modal logic lacks operators to describe the change, origin, and termination of lower-order relations. Although extended dynamic and temporal logics have been applied to Hohfeldian powers (see Section 2), application of these logics to physical laws has been limited. Notably, Baltag and Smets (2005, 2008, 2011, 2012) make a significant contribution by applying propositional dynamic logic to quantum mechanics to provide a more comprehensive description of the nature of quantum states and measurement. Specifically, they argue that by incorporating a logic of action derived from propositional dynamic logic, the "non-classical features of quantum behavior" become "a consequence of the non-classical 'dynamics' of quantum information" (Baltag and Smets, 2012). On their approach, a quantum "system is completely characterized by its potential behavior under all possible interactions." Furthermore, it is the "non-classical nature of quantum-information-extracting actions," inherent in the process of quantum measurement, "that explains the strangeness of quantum behavior" (Baltag and Smets, 2012).

Yet, Baltag and Smets (2005, 2008, 2011, 2012) and related works fail to appreciate the distinction between the first- and second-order nature of states and dynamics in quantum mechanics (again, as the term "second-order" is used here, i.e., not in the ordinary sense of first- and second-order in logic, see Section 2). Indeed, Baltag and Smets (2005, 2008, 2011, 2012) treat both unitary evolution and measurement as similar dynamic processes, failing to sufficiently distinguish how the dynamic nature of each process differs. As I explain in this section, unitary evolution can be suitably described by first-order modal logic without resort to second-order operators. It is only the process of quantum measurement that requires a dynamic, higher-order (in the Hohfeldian sense) logic that allows for the transition between probabilistic and definite states.

The remainder of this Section is divided into two parts: first, a "classical," then a "quantum," ordering of physical law and related states. In classical systems, the nature and evolution of first-order systems are ontologically determinate. Although first-order laws could hypothetically change in a generalized first-order physical system, they would do so in a wholly determinate manner or would maintain their determinate nature following second-order changes to the first-order laws. In quantum systems, the nature and evolution of first-order systems are ontologically indeterminate, in the sense that first-order processes—at an ontological level—do not uniquely determine the outcome of measurement, even absent epistemic indeterminacy. Indeed, in quantum systems, it is necessary for a second-order process to "collapse" ontologically probabilistic states into definite, "classical" (from a Hohfeldian perspective) measurement outcomes.

3.1 The Order of Classical Physical Systems

The key aspect of classical physical law as applied to physical systems is complete determinism, at least with respect to first-order laws that remain unchanged. In other words, if one assumes that first-order laws are eternal, then the only indeterminacy present in physical law is due to the lack of knowledge on the part of any observer rather than any inherent indeterminacy in the laws or systems themselves. In other words, as discussed earlier, classical systems—at least at a first-order level—only exhibit epistemic indeterminacy but not ontological indeterminacy.

Here, I illustrate this notion first by a simple variation of cellular automata (cf. Toffoli, 1977), then by a more realistic physical example (Sichelman, 2024). Notably, the discussion here does not contend that cellular automata can fully model physical systems, but rather that cellular automata share many features that are similar or analogous to features in physical systems.

### 3.1.1 Classical Cellular Automata

Assume a 2 x 2 grid as in Figure 1 ("space") labeled zones A to D, where each square in the grid is occupied by a circle that is either white or black at a given discrete tick of a counter ("time") based upon a set of fixed rules.

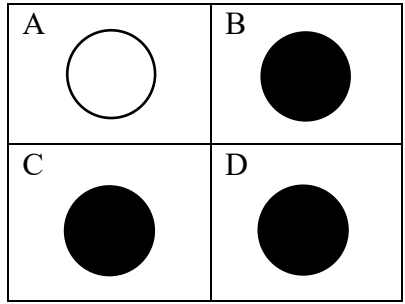

Figure 1. Simplified Cellular Automata.

At t = 0, suppose the circle in cell A is white but cells B, C, and D are black ("initial conditions"). Now suppose only two deterministic rules govern the system:

- Rule 1: For cells A and C, on the next tick, the colors must match whatever the color of the circle was in cells B and D in the previous tick, respectively.
- Rule 2: For cells B and D, on the next tick, the colors must be the opposite whatever the color of the circle was in cells A and C in the previous tick, respectively.

These rules are akin to Hohfeldian obligations in the sense that the physical system must obey Rules 1 and 2. In modal terms, as time progresses (t=1, t=2, etc.), the system necessarily evolves according to these rules. The evolution is completely ontologically determinate and thus classical (both in the physical and in the Hohfeldian sense). Figure 2 illustrates the evolution of the system at two subsequent ticks (t=1, t=2), showing how the colors in each cell change according to Rules 1 and 2.

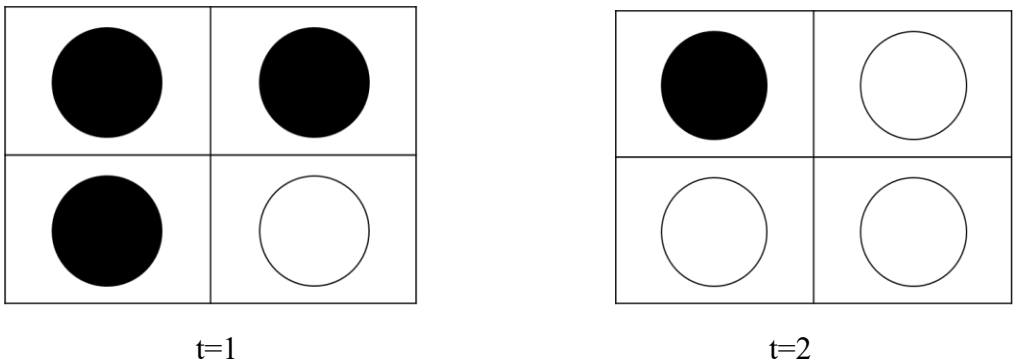

t=1               t=2

Figure 2. Evolution of the Automata for Two Ticks

In a first-order system with unchanging rules, there is no need for resort to second-order rules in order to describe the evolution of the physical system. The only second-order question is the origin of the rules that govern the system, but such questions can be cordoned off as "metaphysical speculation," especially because they do not impact the determinism of the system's evolution within the framework of the given rules. Indeed, even if the rules change, as long as the change is determinate, there is still no need to resort to second-order rules. For instance, suppose it is known in advance that at t=100 that Rule 1 and Rule 2 "flip" so that Rule 1 now requires the colors be opposite and Rule 2 now requires that the colors match. Such a transition can still be described by first-order rules. Even though in this instance a temporal logic may be required to enhance traditional modal or deontic logics, which assume static rule sets, there is no need to resort to second-order rules (in the Hohfeldian sense), as the system remains fully deterministic and can be described within the framework of first-order laws. In other words, even if the deterministic change in rules requires a "meta"-rule to describe the temporal evolution of the rules, the meta-rule is not a second-order physical process akin to a Hohfeldian second-order power that unpredictably intervenes to alter a first-order rule. Rather, the evolution is completely specified via first-order rules and ordinary logic in a wholly deterministic fashion at t=0.

### 3.1.2 Classical Physical Systems

In general, all classical physical systems will, from a Hohfeldian perspective, evolve in a similar fashion to cellular automata governed by static rules (or, again, even dynamic rules that are predictable in advance). In essence, there will be initial conditions, space, time, and physical systems that must (deontic) necessarily (modal) evolve according to the rules (physical laws) in an ontologically determinate fashion.

For instance, consider the motion of a charged particle in an electromagnetic field. With some reasonable simplifying assumptions, the particle's motion is determined completely by the initial conditions of the physical system (the particle's position, velocity, mass, and charge and the strength and direction of electromagnetic field at t=0) and the Newtonian-Einsteinian-Maxwellian laws. In other words, like the cellular automata in the earlier example, the initial conditions plus relevant *first-order* laws supply a complete description of how the system will evolve in spacetime. Furthermore, again like the cellular automata, if the laws are eternal (or even changing in some predetermined fashion), there is no need to resort to second-order rules for a "complete scientific" description of the physical laws.

Indeed, one can construct a Hohfeldian first-order proposition to describe the application of first-order physical laws to the motion of the charged particle (Sichelman, 2024). Specifically, if we postulate a jural proposition, $J_1$, that describes the evolution of the particle's position and momentum from t=0 ($t_0$) to t=1 ($t_1$) and assume it is the universe-at-large that holds the "right" that particle P move in such a manner, we can construct the proposition, $J_1$ as in (16). In the following statement, $J_1(t_0\text{-}t_1)$ is a first-order legal relation that applies from the time $t_0$ to the time $t_1$. The two "legal" actors are $U$, the "unongiverse-at-large," and $P$, the electron. $U$ holds a first-order Hohfeldian *strict-right* vis-à-vis $P$ that the state of affairs in statement 4 occurs. Specifically:

$J_1(t_0-t_1) = U_{r1}P$

1 (specifies the present state of the world). P is an electron that is initially situated at $x_0$, $y_0$, $z_0$ with an initial velocity $v_{x0}$, $v_{y0}$, $v_{z0}$, at time, $t_0$;

2 (further specifies the present state of the world). There is a charge density and current density, $\rho$ and **J**, which generate electric (**E**) and magnetic (**B**) fields;

3 (simplifying assumptions about the present state of the world). Only **E** and **B** affect the motion of the electron, *P*, and the motion of *P* does not affect **E** and **B**;

4 (future state of the world). *P* follows a unique path in spacetime determined by conditions 1-3 and classical "laws" of motion determined by Maxwell's equations and Newton's laws (or Einsteinian laws, if we wish to apply special relativity)) (16)

In other words, given existing conditions (1 and 2) and simplifying assumptions (3) regarding the current state of the world, P is obligated to perform (4). Alternatively, one can use the modal language of necessity rather than the deontic logic of obligation. In either case, the particle must adhere to (4) by determinately following a unique path in spacetime.

### 3.2 The Order of Quantum Physical Systems

Recall that the key difference between classical and post-classical legal systems is that any indeterminacy in a classical system is merely epistemic in nature, whereas post-classical systems admit of ontological indeterminacy—namely, indeterminacy that is inherent in the nature of the system, rather than merely a product of the lack of knowledge on the part of observers. Consistent with many interpretations of quantum mechanics, including the Copenhagen interpretation, one can reasonably assume that the transition from classical to post-classical laws of physics also involves a shift from purely epistemic indeterminacy to systems that admit of ontological determinacy, and then examine the consequences of such an assumption. Similar to the discussion of classical systems, I begin with quantum cellular automata, then proceed to quantum electrodynamics.

#### 3.2.1 The Order of Quantum Cellular Automata

In order to construct quantum cellular automata, at least for illustrative purposes, one can introduce ontological indeterminacy by placing the state each cell into a superposition of possible states, which is resolved only upon "measurement."

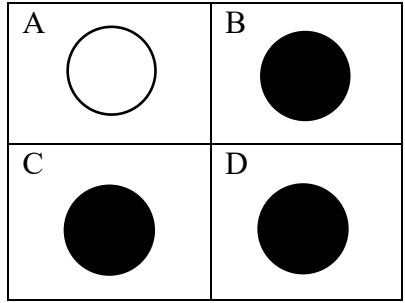

Figure 3. Simplified Quantum Cellular Automata.

Like the prior classical example, at t=0, Figure 3 represents the initial conditions of the system, in which the circle in cell A is white whereas the circles in cells B, C, and D are black. In other words, at t=0, the system is an entirely classical state. Suppose three rules govern the system:

- <u>Rule 1Q</u>: For cell A, on the next tick, the state of the system enters a superposition of 50% of the color previously in cell A and 50% of the color previously in cell B. On the next tick, cell C is in a superposition of 50% of the previous color of cell C and 50% of the previous color of cell D.
- <u>Rule 2Q</u>: For cell B, on the next tick, the state of the system is now in a superposition of 50% of the opposite of the color previously in cell A and 50% of the opposite of the color previously in cell B. The same rule applies for cell D, except that cell D is in a 50/50 superposition of the opposite of the previous colors of cells C and D.
- <u>Rule 3Q</u>: As specified, Rules 1Q and 2Q are applied at each new tick. Immediately after their application, but before the next tick, any cell in a superposition undergoes "measurement," collapsing into either white or black based on the probabilities set by Rules 1Q or 2Q. Thus, while cells briefly enter a superposition state, each cell remains either black or white for the duration of any given tick.

If we use a shade of gray to represent a cell's superposition, we can describe the evolution of the system. Specifically, at t=1, using the above rules, cells A and B are initially in a 50/50 black-white superposition; cell C is black; and cell D is white. Although the ontological superposition of states is not classical, so far, the evolution itself is entirely classical (in the Hohfeldian sense), because it is wholly determined by the initial conditions and rules (specifically Rules 1Q and 2Q). Note that because a superposition is never directly measurable, the "gray" states need not be ontologically realizable, but merely need to represent mathematically the probability of finding the cell in a black or white state upon measurement. See Figure 4.

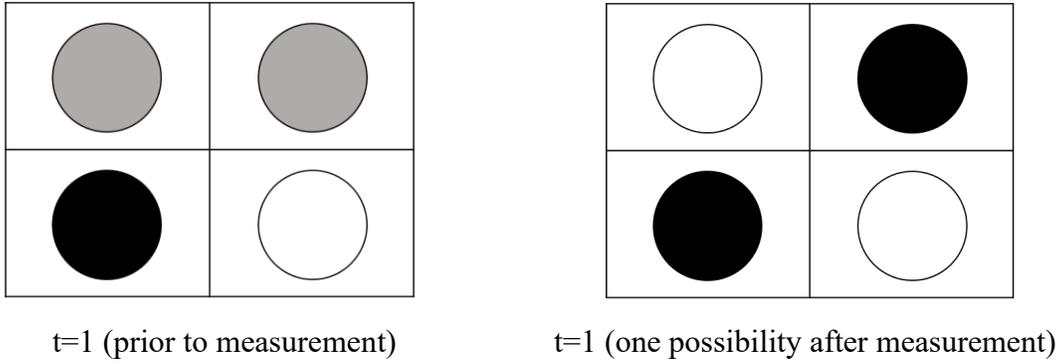

t=1 (prior to measurement)        t=1 (one possibility after measurement)

Figure 4. Evolution of the Quantum Automata

Importantly, there is ontological indeterminacy regarding the state of the system upon measurement. In other words, even with complete information about the system, there is no way to determine the measurement outcome with certainty. Specifically, both cell A and cell B have a 50% probability of being measured white and a 50% probability of being measured black. This makes for different possible outcomes of measurement for cells A and B, respectively: white-black (depicted in Figure 4), white-white, black-white, and black-black.

Unlike the classical case, the precise state of the quantum cellular automata upon measurement will be unpredictable. Notably, these variants are not—by definition in this example—caused by anything within the system per se, nor in the nature of space or time in which the system resides. Rather, from the perspective of the system, they are entirely random. Yet, each of the four outcomes here could be characterized as the implementation of a short-term law from t=0 to t=1. If we consider only the fully measured state at t=1, and that state is black-white (as shown in Figure 4), then the evolution can be depicted as proceeding according to the short-term rule illustrated in Figure 5.

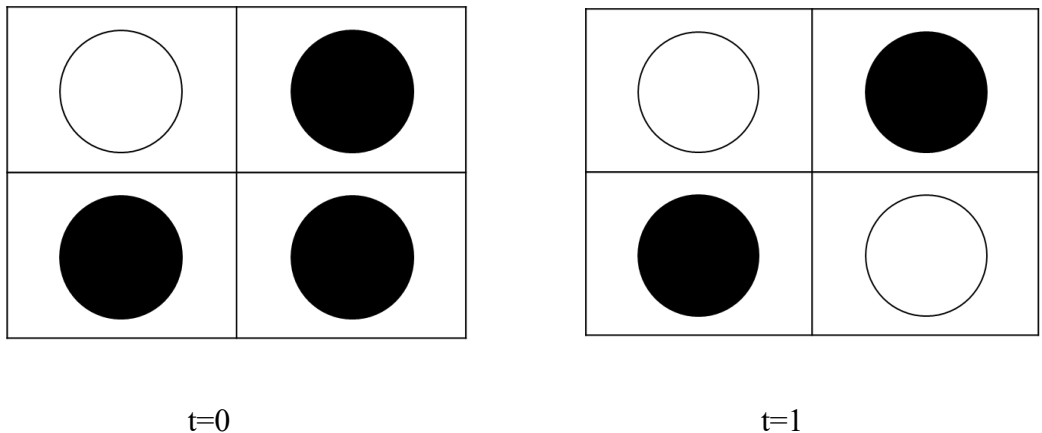

t=0                               t=1

Figure 5. Evolution of the Quantum Automata for Two Ticks (measurement case #1)

We can characterize the evolution of quantum automata using the Hohfeldian categorization of first- *and* second-order laws. As noted earlier, Rules 1Q and 2Q, even though they result in quantum superpositions for each cell, *prior to measurement*, they result in a state evolution that is wholly determinate and thus classical from a Hohfeldian perspective. In other words, while the superposed states themselves are inherently non-classical, their evolution prior to measurement is classical in the sense that one can precisely determine the specific evolution of the initial states according to Rules 1Q and 2Q.

Yet, as noted neither Rules 1Q and 2Q, nor Rule 3Q, can predict the result of measurement in Figure 5—in other words, what colors appear in cells A and B at t=1. Rather, only laws or rules outside of the system could possibly explain what results when a measurement is made. In critical part, these laws must be second-order in the Hohfeldian sense. The key insight is that for each of the four possible evolutions from t=0 to the measured states in t=1, one could add a new rule 4Q that dictates the outcome of measurement. For instance, Rule 4Q could read, "Upon measurement, a random number generator outside of the (first-order) system is used to choose black or white for each cell in a superposition in accordance with the underlying probabilities of the quantum state of the system."

In essence, the measurement process is a second-order process, because it effectively generates a classical "microlaw"—again a law that applies to a given system rather than globally—that dictates how the classical state (in the Hohfeldian sense) of the system at t=0 evolves to the classical state (in the Hohfeldian sense) of the system at t=1. In other words, if one only observed the outcome in Figure 5, one could construct a microlaw to explain the evolution of the system from t=0 to t=1. There are four such microlaws, corresponding to each possible outcome of measurement. Such selection is analogous to a judge determining whether a litigant breached an obligation in a case in which the legal relation is ontologically indeterminate until final judgment.

In sum, at least based on the assumptions made here for cellular automata, if we believe that the outcome of a quantum measurement has a physical cause, then such a cause must emanate, at least in part, from a second-order physical process. Before turning to this discussion further in Section 4, it is useful to consider the nature of ontological indeterminacy and measurement in the context of an actual physical system, again, a charged particle in an electromagnetic field.

### 3.2.2 The Order of Quantum Electrodynamics

A charged particle's non-relativistic motion through an electromagnetic field can be described by quantum electrodynamics, which incorporates Feynman's "sum over histories" approach (Feynman and Hibbs, 1965). In contrast to the classical case, in which a particle determinately travels along one unique path from $(x_0, t_0)$ to $(x_1, t_1)$, in quantum theory, a particle whose wave function has not yet been measured may be viewed as effectively traveling along *every possible path* from $(x_0, t_0)$ to $(x_1, t_1)$. For simplicity, in Figure 6 below, the particle may effectively travel on only four paths in spacetime from $(x_0, t_0)$ to $(x_1, t_1)$.

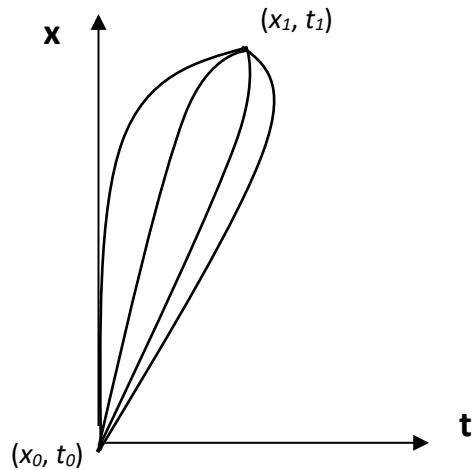

Fig. 6. The Quantum Motion of a
Particle in Two-Dimensional Spacetime

If we let $(x_1, t_1)$ be any general point in spacetime $(x, t)$, the Feynman "sum over histories" approach provides a method of calculating a sum of weighted "contributions" from each path to the evolution in spacetime of a quantum state $|\psi\rangle$ of a particle $P$ in spacetime to a given endpoint $(x_1, t_1)$.[12] More specifically, in this approach, the paths "interfere" with another constructively and destructively, resulting in a complex calculation that provides the probability of a particle reaching a particular endpoint, $(x_1, t_1)$, but no clear probability that a particle took any particular path.

Suppose we allow the particle's location to evolve in time without any measurement. Notably, like the situation of the evolving quantum cellular automata prior to measurement, in the absence of any attempt to determine where the particle is located—the evolution of a quantum state, $|\psi\rangle$, of a particle, $P$, is *from a Hohfeldian standpoint* entirely *classical*. In quantum electrodynamics, one can specify a *deterministic* unitary evolution operator, $U(x, t)$, which precisely describes how a particle's quantum state, $|\psi\rangle$, evolves in spacetime (Adler, 2003).

If we attempt to measure which path the particle en route to its destination, then the situation is no longer determinate and, from a Hohfeldian perspective, is post-classical. In our idealized example, we can attempt to determine the specific path a particle takes by placing measuring devices in the middle of each of the four paths, and when a position measurement is made in this instance, only one of the four paths will register (Mensky, 1993). Quantum measurement—like a legal judgment—results in a classical outcome (in the Hohfeldian sense) and thereby eliminates alternative historical states of evolution of the particle (Zurek, 2003).

---

[12] Ultimately, quantum field theory treats particles as the emanations of underlying fields, but for simplicity, I refer to "the particle" in the discussion here. For the conceptual difficulties involved in classifying "particles" as basic objects in quantum field theory, see Kuhlman (2023).

The elimination of possible historical states of evolution is analogous to the situation we encountered for quantum cellular automata. In essence, measurement selects a specific microlaw that explains the transition from one measured state, here, $(x_0, t_0)$, to another measured state, here, $(x_1, t_1)$. In other words, in the event one measures a specific path the particle is traveling along, one can no longer view the particle as subject to *one, classical-like* unitary evolution. Rather, akin to the four possible outcomes of the measurement of the cellular automata in the example above—in the present example, one may posit *four separate classical-like legal relations* that can be used to describe the four possible results (i.e., four particular paths) of quantum measurement. Thus, like a legal actor who may in essence be subject to different laws depending on the result of a judicial ruling, the particle here is potentially subject to four separate first-order legal relations—i.e., four separate classical-like laws of nature.[13] In a Hohfeldian sense, it is as if each potential path of the particle from $(x_0, t_0)$ to $(x_1, t_1)$ represents a different classical-like law that may be instantiated for the particle $P$ depending on the result of a measurement of the particle's particular path. When a suitable measurement selects only one path, the selected path becomes a *positive duty* (deontic)/necessary (modal) relation and the unselected paths become *negative duty* (deontic)/impossible (modal) relations.

And from the earlier discussion, we know—at least in the Hohfeldian sense—how laws are chosen: via higher-order powers.[14] Specifically, a second-order Hohfeldian power alters the probabilities associated with each potential outcome (i.e., the probability of each path being taken) so that only one path remains with a 100% probability. Thus, like a judge who decides whether a plaintiff is subject to a given law or not via a second-order power, if there is a causal story to be told, *a quantum measurement executes the analogue of a second-order Hohfeldian power to collapse the wave function $|\psi\rangle$ of a particle P*.[15] More specifically, this collapse selects one of many competing classical-like states—in effect, laws—that the particle could have "obeyed" before the measurement. Importantly, unlike classical physics, in which second-order laws could be relegated to the realm of metaphysics, second-order laws appear essential to provide a causal, physical explanation in quantum physics.

Positing that quantum measurement is a second-order physical process raises several important questions. First, if the process does not depend on the ordinary first-order laws, how do the ordinary fundamental constituents (matter, fields, spacetime) effectuate a second-order process? Alternatively, if the ordinary fundamental constituents do not effectuate a second-order process on their own, what other constituents are necessary? If additional constituents are involved, do they occupy the same spacetime as ordinary fundamental constituents? If not, where do they reside? Second, what specifically triggers the second-order process of quantum measurement? In other words, when does measurement occur, and when does it not? Third, how does the approach offered here compare to other theories of quantum measurement?

---

[13] These paths are "classical-like" in the sense that they select a single path through spacetime at the time of measurement, but notably differ from the actual classical laws, which dicate a single path regardless of measurement (Galiev, 2020).

[14] In essence, the second-order measurement power in quantum mechanics will (nearly) instantaneously convert a probabilistic superposition of states, $|\psi\rangle$, into some determinate eigenstate of $|\psi\rangle$ that one might expect from a classical measurement (albeit a result that might not follow from the "macroscopic" classical law that is the "large-scale" limit of the quantum formalism) (cf. Kastner et al., 2016).

[15] In essence, the only means by which a true (i.e., quantum-like) superposition of possible first-order states of any system can be reduced to a single, first-order state is via a second-order process. This approach is in contrast to the mere absence of knowledge of which specific first-order state a system occupies, which in turn is revealed by a classical measurement via a first-order physical process. It also differs from theories that explain measurement via first-order physical processes that transform the state of the quantum system.

These critical questions are explored in future work (Sichelman, 2025). Briefly, I propose that the level of quantum decoherence of a quantum system is directly proportional to the probability of triggering a second-order physical process that selects from among the potential classical-like states of—in essence, microlaws governing—the system. Specifically, even in empty space, there is a very small probability of system collapse due to the production of virtual particles interacting with the system. This rate is vanishingly low, so that a "single particle" traveling in empty space nearly universally will obey the Schrodinger equation. However, as more and more particles vigorously "interact" with one another, they will essentially undergo continuous decoherence, leading to a high probability of collapse, leading to a "macroscopic" object that appears "classical." Because measuring devices are typically macroscopic, "classical-like" devices, small numbers of quantum "particles" interacting with such devices will typically decohere sufficiently to "collapse" the wave packet, consistent with the standard interpretation of measurement. On the other hand, the approach here does not necessitate measuring devices or observers to "collapse" the wave packet—rather, ordinary physical processes that trigger second-order processes are sufficient. In this regard, "measurement" in this approach does not require adding terms to standard quantum mechanics, as in spontaneous collapse theories, but emerges because it is triggered by a *second-order* physical process.

## 5 Conclusion

Throughout the history of law, physics has informed the structure of legal systems and associated laws. For example, John Adams applied the concept of equilibrium in classical mechanics to constitutional law, helping to develop the principle of "checks and balances" (Shachtman, 2014), and the drafting of the U.S. Constitution was influenced by Montesquieu's analogies between legal concepts and Newton's laws (Koukoutchos, 1988). Although there are historical instances of physicists drawing upon legal principles for inspiration—for example, Leibniz's conception of physical law (Dong, 2024)—very seldom, particularly in the 20th and 21st centuries, have physicists leveraged legal philosophy to model theory. This article reverses that trend by drawing upon the work of Hohfeld (1913) to describe the "order" of physical law. By offering a more formal link between scientific and legal systems than the prior literature, this article provides a richer notion of physical law with useful applications. For example, the article posits that the concept of second-order law is essential to understanding the nature of quantum measurement, the details of which will be explored in future work (Sichelman, 2025).